\documentclass[aps,prl,twocolumn,showpacs,superscriptaddress]{revtex4}

\newcommand{\extraspace}{\addtolength{\abovedisplayskip}{0mm} 
                        \addtolength{\belowdisplayskip}{0mm} 
                        \addtolength{\abovedisplayshortskip}{0mm} 
                        \addtolength{\belowdisplayshortskip}{0mm}} 
\newcommand{\be}{\begin{equation}\extraspace} 
\newcommand{\ee}{\end{equation}} 
\newcommand{\bea}{\begin{eqnarray}\extraspace} 
\newcommand{\eea}{\end{eqnarray}} 

\newcommand{\wtPsi}{\widetilde{\Psi}}

\newcommand{\nonu}{\nonumber \\[2mm]} 
  
\begin{document}
\title{Separation of spin and charge in paired spin-singlet 
       quantum Hall states}
\author{E.~Ardonne}
\affiliation{
Institute for Theoretical Physics,
Valckenierstraat 65, 1018 XE Amsterdam, THE NETHERLANDS}
\author{F.J.M.~van Lankvelt}
\affiliation{
Institute for Theoretical Physics,
Valckenierstraat 65, 1018 XE Amsterdam, THE NETHERLANDS}
\author{A.W.W.~Ludwig}
\affiliation{
Department of Physics, University of California, Santa Barbara, CA 93106}
\author{K.~Schoutens}
\affiliation{
Institute for Theoretical Physics,
Valckenierstraat 65, 1018 XE Amsterdam, THE NETHERLANDS}

\date{February 5, 2001}

\begin{abstract}
We propose a series of paired spin-singlet quantum Hall states,
which exhibit a separation of spin and charge degrees of freedom.
The fundamental excitations over these states, which have filling 
fraction $\nu=\frac{2}{2m+1}$ with $m$ an odd integer, 
are spinons (spin-$\frac{1}{2}$ and charge zero) or fractional 
holons (charge $\pm \frac{1}{2m+1}$ and spin zero). The braid statistics 
of these excitations are non-abelian. The mechanism for the separation 
of spin and charge in these states is topological: spin and charge 
excitations are liberated by binding to a vortex in a $p$-wave 
pairing condensate. We briefly discuss related, abelian spin-singlet
states and possible transitions.

\end{abstract}


\maketitle

Strongly correlated electrons in low dimensional systems 
are known to exhibit physical phenomena that are surprising 
and, at first sight, counterintuitive. Among these is the 
remarkable phenomenon of {\it quantum number fractionalization}: 
elementary excitations in strongly interacting many-electron 
systems can have quantum numbers (for spin and charge) that 
are fractions of those of the electron. This fractionalization 
can take the form of a separation of spin and charge, or of a
fractionalization of the electric charge of the electron.

In $D\!=\!1$ spatial dimension, the separation of spin and charge is 
well understood. It is seen in explicit solutions of specific
integrable model systems (Hubbard and supersymmetric $t$-$J$ 
models). The general framework of the Luttinger Liquid has made
it clear that in 1+1 dimensions the separation of spin and charge 
is a generic feature, which does not require any fine tuning of 
the interactions among the electrons.

In spatial dimensions $D\!=\!2$ or higher, spin and charge tend to 
confine and a separation of the two is only possible under very 
special conditions. It has been proposed that the key feature
underlying the anomalous behavior of the cuprate high-$T_c$
materials is precisely a separation of spin and charge \cite{And}, 
and concrete scenarios, based on ${\bf Z}_2$ or $U(1)$ gauge theories,
have been put forward \cite{spch}. 

In this paper, we focus on the quantum Hall (qH) regime, which
is relevant for $2D$ electrons in strong magnetic fields, and for 
rotating Bose-Einstein condensates \cite{WG}.
In particular, we discuss the separation of spin and charge in 
the qH regime. Specifically, we propose a series of paired spin-singlet 
qH states, of filling fraction $\nu=\frac{2}{2m+1}$,
which are generalizations of the Moore-Read or pfaffian states for 
spin polarized electrons. The fundamental excitations over these
states are spinons (with spin $\frac{1}{2}$ and zero charge) and 
holons (with zero spin and fractional charge $\pm \frac{1}{2m+1}$,
in units of the charge of the electron). The braid statistics of 
these excitations are non-abelian, and thereby the paired 
spin-singlet states fall in the category of `non-abelian qH states'.

It is important to stress that the more conventional `abelian'
spin-singlet qH states (such as the Halperin states with label 
$(m+1,m+1,m)$, see below) do not exhibit a separation of spin and 
charge. The excitations over such states are conveniently 
analyzed in terms of a `spin-charge decomposition' \cite{MR,BF,MiR} 
but this is subject to certain gluing conditions (expressing
locality of the excitation w.r.t.\ the electrons),
which exclude single spinons or holons from the (bulk)
physical spectrum. 
The essential feature that liberates spin and charge in the paired 
states proposed here 
is the presence of the pairing condensate: by binding to a 
vortex in the pairing condensate, the spin and charge 
excitations become local with respect to the electrons in the 
ground state, and they can propagate independently.

It is illustrative to compare the separation of spin and charge
in the paired spin-singlet states with the fractionalization of 
charge in paired, so-called $q$-pfaffian, spin polarized states. 
For the $q$-pfaffian states, Laughlin's
gauge argument gives that the adiabatic insertion of a single
flux quantum will produce an excitation of charge $\frac{1}{q}$.
However, as in the case of BCS superconductors, the presence of 
the pairing condensate leads to a reduction of the elementary
flux quantum by a factor of $2$, and thereby the unit-flux Laughlin 
quasi-particles are separated into two constituents each carrying 
a charge $\frac{1}{2q}$. In a similar way, conventional quasiparticles 
(carrying spin and charge) over a paired spin-singlet state are 
separated into spinons and holons.

Before we present the paired spin-singlet states, we briefly recall 
some facts about spin-singlet states and paired states in the qH regime.
Despite the presence of strong magnetic fields in the qH
regime, there is experimental motivation to study states that
are not (fully) spin polarized (see e.g. \cite{Eis}). In many qH
systems, the energy scale for the Zeeman splitting is relatively 
low, and it can be further suppressed by the application of hydrostatic
pressure. Using this technique, combined with a tilted field technique,
spin transitions in the qH regime can be studied \cite{Kang}. 
The simplest qH states that are singlets w.r.t. the $SU(2)$ spin
symmetry are the Halperin states \cite{Halp}
\bea
\label{eq:halp}
\lefteqn{ \wtPsi_{\rm  SS}^{(m+1,m+1,m)}
(z^\uparrow_1,\ldots,z^\uparrow_N ; z^\downarrow_1,\ldots,z^\downarrow_N) =}
\\[2mm] && \quad
\Pi_{i<j}(z^\uparrow_i-z^\uparrow_j)^{m+1} 
\Pi_{i<j}(z^\downarrow_i-z^\downarrow_j)^{m+1} 
\Pi_{i,j}(z^\uparrow_i-z^\downarrow_j)^{m}, 
\nonumber
\eea
where $z^\uparrow_i$ and $z^\downarrow_i$ 
are the coordinates of the spin up
and spin down electrons, respectively, and $m$ is an even
integer. The state Eq.~(\ref{eq:halp}) has filling fraction 
$\nu=2/(2m+1)$. 
Here and below we display reduced qH wave functions 
$\wtPsi(x)$, which are related to the actual wave functions 
$\Psi(x)$ via $\Psi(x) = \wtPsi(x) \exp{(-\sum_i 
\frac{|x_i|^2}{4 l^2})}$ with $x_i=z^\uparrow_i,
z^\downarrow_i$ and 
$l=\sqrt{\frac{\hbar c}{e B}}$ the magnetic length. 
Hierarchies of more general (abelian) 
spin-singlet states were studied in \cite{Rez,WDJ,LF,MiR}. 

In a 1991 paper, Moore and Read introduced the notion of 
a paired qH state and discussed the so-called
$q$-pfaffian states at filling $\nu=\frac{1}{q}$ (with
$q$ even) \cite{MR}. It is believed that this state (with $q=2$) is 
at the origin of the observed qH plateau at filling 
fraction $\nu=\frac{5}{2}$ (see \cite{Read} for a recent
review). The wave function for the 
$q$-pfaffian is given by
\be
\wtPsi^{(q)}_{\rm pf}(z_1,\ldots,z_N) =
{\rm Pf} \left( \frac{1}{z_i-z_j} \right)
\prod_{i<j} (z_i-z_j)^q \ ,
\label{eq:qpfaffian}
\ee
where the pfaffian factor for an antisymmetric matrix
$M_{ij}$ is defined as ${\rm Pf}(M_{ij})= 
\mathcal{A}\prod_{i\ {\rm even}} M_{i-1,i}$, with 
$\mathcal{A}$ denoting anti-symmetrization.
In \cite{ARRS}, the pfaffian states were generalized
to a series of non-abelian spin-singlet (NASS) states, at filling 
$\nu=\frac{4}{4M+3}$ with $M$ an odd integer. These states
exhibit a pairing of like spins. The excitations over these 
NASS states have non-abelian statistics, but there is no 
separation of spin and charge.

In the paired spin-singlet states that we propose here, 
the pairing takes place in the charge sector, irrespective 
of the spin of the electrons. This leads to a wave function
\bea
\lefteqn{ \wtPsi_{\rm paired}^{(m)}
(z^\uparrow_1,\ldots,z^\uparrow_N;z^\downarrow_1,\ldots,z^\downarrow_N) =}
\nonu && \quad
{\rm Pf}\left( \frac{1}{x_i-x_j} \right)
\wtPsi_{\rm  SS}^{(m+1,m+1,m)}(z^\uparrow_i;z^\downarrow_j) \ ,
\label{eq:paired}
\eea
where $x_l=z^\uparrow_i,z^\downarrow_j$, $m$ is now an odd integer 
and the filling fraction is $\nu=\frac{2}{2m+1}$.
There exists a hamiltonian for which this state is the unique ground state
\cite{Rezpri}. One way to study the excitations over this state is by using 
this hamiltonian. Here we will proceed by analyzing the state 
Eq.~(\ref{eq:paired}) and its excitations using an associated conformal field 
theory (CFT).

\begin{figure}[ht]
\setlength{\unitlength}{.8 cm}
\begin{picture}(6,6)(-.5,-.5)
\put(-.5,-.5){\line(1,1){6}}
\put(-.5,5.5){\line(1,-1){6}}
\put(0,0){\circle*{.2}}
\put(0,2.5){\circle*{.2}}
\put(0,5){\circle*{.2}}
\put(2.5,0){\circle*{.2}}
\put(2.5,5){\circle*{.2}}
\put(5,0){\circle*{.2}}
\put(5,2.5){\circle*{.2}}
\put(5,5){\circle*{.2}}
\put(1.25,1.25){\circle{.2}}
\put(1.25,3.75){\circle{.2}}
\put(3.75,1.25){\circle{.2}}
\put(3.75,3.75){\circle{.2}}
\put(5.2,4.7){\vector(1,1){.5}}
\put(5.2,0.3){\vector(1,-1){.5}}
\put(5.8,5.2){c}
\put(5.8,-.4){s}
\put(5.2,2.5){$\Psi^\downarrow$}
\put(-.7,2.5){$\overline{\Psi}^\uparrow$}
\put(2.5,5.2){$\Psi^\uparrow$}
\put(2.5,-.5){$\overline{\Psi}^\downarrow$}
\put(.1,5.2){$\Delta^{\uparrow\uparrow}_s$}
\put(4.5,-.5){$\Delta^{\downarrow\downarrow}_s$}
\put(4.5,5.2){$\Delta_c$}
\put(.1,-.5){$\overline{\Delta}_c$}
\put(4,3.4){$\phi_c$}
\put(.7,1.4){$\overline{\phi}_c$}
\put(.7,3.4){$\phi^\uparrow_s$}
\put(4,1.4){$\phi^\downarrow_s$}
\end{picture}

\caption{Roots and weights of the algebra $SO(5)$.
The condensate operators $\Psi$ and $\Delta$ are
associated to the eight roots (filled symbols)
and the fundamental excitations $\phi_{s,c}$
correspond to the weights of the spinor representation 
(open symbols).}

\end{figure}
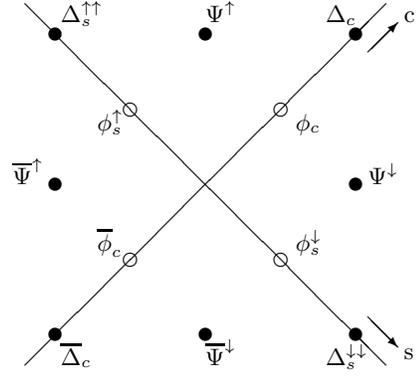

Following the CFT-qH correspondence outlined in 
\cite{MR}, one quickly finds that the CFT associated
to the (bosonic) paired spin-singlet state at $m=0$ 
is the (chiral) CFT based on the affine Kac-Moody algebra 
$SO(5)_1$. For this algebra, the eight currents associated 
to the roots of $SO(5)$ can be written in terms of spin 
and charge bosons $\varphi_{s,c}$ and a Majorana fermion 
$\psi$.
[The assignment of spin and charge quantum numbers to 
the weights and roots of $SO(5)$ is indicated
in fig.~1.] For general $m$, the `condensate' operators 
$\Psi$ and $\Delta$ are obtained from these currents by 
the substitution $\varphi_c \to \sqrt{2m+1}\, \varphi_c$,
\bea && \hspace{-2mm}
\Psi^{\alpha} = \psi \, e^{i \sqrt{\frac{2m+1}{2}} \varphi_c 
         \pm \frac{i}{\sqrt{2}}  \varphi_s} , \ 
\overline{\Psi}^{\alpha} = \psi \, e^{-i \sqrt{\frac{2m+1}{2}} \varphi_c 
         \pm \frac{i}{\sqrt{2}}  \varphi_s} 
\nonu && \hspace{-2mm}
\Delta_c = e^{i\sqrt{4m+2}\, \varphi_c} , \  
\overline{\Delta}_c = e^{-i\sqrt{4m+2}\, \varphi_c} , \ 
\Delta_s^{A} = e^{\pm i\sqrt{2} \, \varphi_s} , 
\nonumber \\ &&
\label{eq:so5currents}
\eea
with $\alpha=\uparrow,\downarrow$ referring to the
spin eigenvalue $s_z=\pm \frac{1}{2}$ and 
$A=\uparrow\uparrow,\downarrow\downarrow$. 
The quantum numbers
$q$ (charge) and $s_z$ are measured by the operators
$Q = -i \sqrt{\frac{2}{2m+1}}\oint \frac{dz}{2\pi i} \partial 
\varphi_c$ and
$S_z = \frac{i}{\sqrt{2}}\oint \frac{dz}{2\pi i} \partial 
\varphi_s$.
The wave function Eq.~(\ref{eq:paired}) is obtained as a correlator
of $N$ spin-up electrons $\Psi^{\uparrow}$ and $N$ spin-down 
electrons $\Psi^{\downarrow}$, together with a neutralizing 
background charge. 
The CFT description makes it easy to identify the 
fundamental (quasi-particle) excitations.
For $m=0$ they are the operators that generate the spinor 
(4-dimensional) representation of the $SO(5)_1$ current algebra.
For general $m$ these become
\be
 \phi_c = \sigma \, e^{\frac{i}{\sqrt{4m+2}} \varphi_c}, 
\
 \overline{\phi}_c = \sigma \, e^{-\frac{i}{\sqrt{4m+2}} \varphi_c}, 
\
 \phi_s^{\alpha} = \sigma \, e^{\pm \frac{i}{\sqrt{2}} \varphi_s},
\label{eq:fundex}
\ee
where $\sigma(z)$ is the so-called spin field associated to
the Majorana (Ising) fermion $\psi(z)$. Higher excitations, such 
as those constituting the vector representation, 
can be generated by bringing together two or more of the 
fundamental excitations. The expressions Eq.~(\ref{eq:fundex}) show 
that the fundamental excitations can be characterized as spinons 
$\phi_s^{\alpha}$ (spin-$\frac{1}{2}$ but 
no charge) and holons $\phi_c$, $\overline{\phi}_c$ (of charge 
$\pm \frac{1}{2m+1}$ and zero spin).

To illustrate the separation of spin and charge, we present explicit
wave functions for excited states. We first consider an
abelian excitation, with spin down ($s_z=-\frac{1}{2}$) and
charge $\frac{1}{2m+1}$, at location $w$. Its wave function
takes the familiar form
\be
\prod_i(z_i^\uparrow-w) \, \wtPsi_{\rm paired}^{(m)}\ .
\label{eq:exabelian}
\ee
The important observation is now that, starting from this 
wave function, one can separate the locations of the spin 
and charge parts of this excitation, creating a spinon 
at position $w_s$ and a holon at $w_c$. In the corresponding 
wave function, the pfaffian factor in Eq.~(\ref{eq:paired}) 
is replaced by (compare with \cite{MR})
\begin{equation}
{\rm Pf}\left(\frac{\Phi(x_i,x_j;w_c,w_s)}{x_i-x_j}\right)
	\prod_i(x_i-w_c)^{1/2}
	\frac{\prod_i(z_i^\uparrow-w_s)^{1/2}}
             {\prod_j(z_j^\downarrow-w_s)^{1/2}},
\label{eq:excited}
\end{equation}
where
\begin{equation}
\Phi(x_i,x_j;w_c,w_s)=\left(
	\frac{x_i-w_c}{x_j-w_c}
	\frac{x_j-w_s}{x_i-w_s}
	\right)^{1/2}
+ i \leftrightarrow j.
\end{equation}
That (\ref{eq:excited}) in fact defines a well-behaved electronic wave
function can be seen by noting that it is identical to
\begin{equation}
\frac{1}{\prod_j(z_j^\downarrow-w_s)}{\rm Pf}\left(
	\frac{(x_i-w_c)(x_j-w_s)+i\leftrightarrow j}{x_i-x_j}\right) .
\end{equation}
In the limit where $w_s,w_c \to w$, spin and charge recombine and
the wave function reduces to Eq.~(\ref{eq:exabelian}).
Note that the factor $\prod_j(z_j^\downarrow-w_s)^{-1}$ should 
be regularized and projected onto the lowest Landau level in the same
way as the wave functions for quasi particles over the Laughlin
states \cite{Laugh}.

The charge of the holon excitation equals $\frac{1}{2} \Phi_0 \sigma_H$
(with $\Phi_0={h \over e}$ the flux quantum),
showing that the creation of a single holon involves the
insertion of a half-quantum of magnetic flux, which is the canonical
flux quantum in the presence of a pairing condensate. This
flux insertion is accompanied by a vortex in the pairing condensate,
and this brings in the factor $\sigma(z)$ in the expressions
Eq.~(\ref{eq:fundex}). The role of the vortices in this 
discussion is similar to the role of visons in the
Senthil-Fisher theory \cite{SF}.

An important feature that is implied by the presence of
spin-fields $\sigma(z)$ in the expressions Eq.~(\ref{eq:fundex})
for the spinons and holons, is that the braid
statistics of these excitations will be non-abelian. This 
feature is analogous to the non-abelian statistics of the charge
$\frac{1}{2q}$ excitations over the (spin-polarized)
$q$-pfaffian state, and we refer to the literature for
a discussion \cite{MR,RR,NW}.

It is well-known that the $q$-pfaffian spin-polarized state is 
closely related to two abelian states at filling 
$\nu=\frac{1}{q}$: the two-layer $(q+1,q+1,q-1)$ state and 
a strong pairing state which is a Laughlin state of strongly 
paired electrons. Possible transitions among these three states 
have been discussed in the literature (see e.g. \cite{NW,RG,Wen}).
In the spin-singlet situation, we may similarly identify two
series of abelian spin-singlet states at $\nu=\frac{2}{2m+1}$
that allow for a transition into the pfaffian spin-singlet
state Eq.~(\ref{eq:paired}): a two-layer state associated
to $SO(6)$ and a strong pairing state. The wave function for 
the two-layer state reads
\bea
\label{eq:so6}
&&
\lefteqn{\wtPsi_{\rm 2-layer}^{(m)}
(\{ z^{\uparrow t}_i, z^{\downarrow t}_i, z^{\uparrow b}_i, 
    z^{\downarrow b}_i \}) =}
\\[2mm] && 
\Pi_{i<j}(z^{\uparrow t}_i-z^{\uparrow t}_j)^{m+2} 
\Pi_{i<j}(z^{\downarrow t}_i-z^{\downarrow t}_j)^{m+2}
\nonu && 
\Pi_{i<j}(z^{\uparrow b}_i-z^{\uparrow b}_j)^{m+2} 
\Pi_{i<j}(z^{\downarrow b}_i-z^{\downarrow b}_j)^{m+2} 
\nonu && 
\Pi_{i,j}(z^{\uparrow t}_i-z^{\downarrow t}_j)^{m+1} 
\Pi_{i,j}(z^{\uparrow b}_i-z^{\downarrow b}_j)^{m+1} 
\Pi_{i,j}(z^{\uparrow t}_i-z^{\uparrow b}_j)^{m} 
\nonu && 
\Pi_{i,j}(z^{\downarrow t}_i-z^{\downarrow b}_j)^{m} 
\Pi_{i,j}(z^{\uparrow t}_i-z^{\downarrow b}_j)^{m-1} 
\Pi_{i,j}(z^{\downarrow t}_i-z^{\uparrow b}_j)^{m-1}
\nonumber
\eea
where the indices $t,b$ refer to the top and bottom
layers. This wave function arises as a correlator of 
two-layer spinful electron operators which, in the
case $m=0$, generate an $SO(6)_1$ affine Kac-Moody
algebra.

The strong pairing state is an 
abelian state of strongly bound pairs with quantum numbers 
$(q\! =\! -2,s_z\! =\! 0)$ and $(q\! =\! 0,s_z\! =\! 1)$, which are 
the operators
$\Delta_c$ and $\Delta_s^{\uparrow\uparrow}$ in fig.~1. 
Spin and charge are decoupled from the start, and 
(putting $m=0$) we can associate to this state the 
symmetry $SO(4) \sim SU(2)_s \times SU(2)_c$.

There are various ways to understand and describe possible 
transitions among the three types of paired spin-singlet states at
$\nu = \frac{2}{2m+1}$. Such transitions are expected when
electrons in the two-layer state are subjected to 
increasing interlayer interactions. A useful framework
is that of $K$-matrices describing the topological order
of the various states \cite{Wen1}. 
[For this discussion, we refer to
the states via their associated $SO(6)$, $SO(5)$ or
$SO(4)$ symmetries.] For the $SO(6)$ states, the naive
$K$-matrix for the four electron operators
$(\uparrow,t)$, $(\downarrow,t)$, $(\uparrow,b)$,
$(\downarrow,b)$ is singular. After a reduction to three
independent condensate operators we find the following
`qH data' 
\bea
\label{eq:so6qHdata}
&& {\bf K}_e = \left( \begin{array}{ccc}
m+2 & m & 2m+1 \\
m & m+2 & 2m+1 \\
2m+1 & 2m+1 & 4m+2 \\
\end{array} \right) 
\\[2mm] &&
{\bf q}_e = -(1,1,2), \quad
{\bf s}_e = (\uparrow,\uparrow,0), \quad
{\bf l}_e = (t,b,\cdot) \ , 
\nonumber
\eea
where ${\bf q}_e$, ${\bf s}_e$ and ${\bf l}_e$ specify 
the charge, spin and layer index for an appropriate basis 
of `electron' operators, which build the qH condensate. 
By applying a duality
transformation (${\bf K}_\phi={\bf K}_e^{-1}$, ${\bf q}_\phi 
= - {\bf K}_\phi {\bf q}_e$, etc.) one obtains the topological 
data for a basis of quasi-hole excitations \cite{Wen1,ABGS}.

Starting from this characterization of the topological
order in the $SO(6)$ state, the topological order of 
the $SO(5)$ and $SO(4)$ states can be obtained in 
a systematic manner \cite{ABGS,Wen}. For the $SO(5)$
state, the resulting description employs a so-called 
pseudo-particle whose role it is to account for the 
degeneracies that are associated to the non-abelian 
braid statistics. Choosing $\Psi^\uparrow$,
$\Delta_s^{\uparrow\uparrow}$ and $\Delta_c$ as the
fundamental condensate operators, we find 
\bea
{\bf K}_e = \left( \begin{array}{ccc}
m+2 & 1 & 2m+1 \\
1 & 2 & 0 \\
2m+1 & 0 & 4m+2 \\
\end{array} \right) 
\ , \ 
\begin{array}{l}
{\bf q}_e = -(1,0,2) \\
{\bf s}_e = (\uparrow,\uparrow\uparrow,0)
\end{array} && \nonu
{\bf K}_\phi = \left( \begin{array}{ccc}
1 & -\frac{1}{2} & -\frac{1}{2} \\
-\frac{1}{2} & \frac{3}{4} & \frac{1}{4} \\
-\frac{1}{2} & \frac{1}{4} & \frac{2m+3}{8m+4} \\
\end{array} \right) 
\ , \ 
\begin{array}{l}
{\bf q}_\phi =  (0,0,\frac{1}{2m+1}) \\
{\bf s}_\phi =  (0,\downarrow,0)
\end{array} &&  
\label{eq:so5qHdata}
\eea
It is the first particle in the $\phi$-sector that is 
interpreted as a pseudo-particle, the other two have
quantum numbers corresponding 
to $\phi_s^\downarrow$ and $\overline{\phi}_c$. The
matrix ${\bf K}_\phi$ is of a general form first proposed
in \cite{GSBCR}; for the interpretation of $K$-matrices for 
non-abelian qH states we refer to \cite{ABGS}. We remark
that the ground state degeneracy on the torus is not
simply given by $|{\rm det}{\bf K}_e|$, as is the case for abelian
qH states; the actual value here is $3(2m+1)$.
 
A further reduction leads to the following qH data for the 
strong pairing $SO(4)$ state (the data for the $\phi$ sector
is obtained by the duality mentioned above)
\be
{\bf K}_e = \left( \begin{array}{cc}
2 & 0 \\
0 & 4m+2 \\
\end{array} \right), \ 
{\bf q}_e = -(0,2), \ 
{\bf s}_e = (\uparrow\uparrow,0) \ .
\ee
This same set of qH data can be obtained by starting from 
the $SO(6)$ data Eq.~(\ref{eq:so6qHdata}) and condensing 
quasiparticle-quasihole pairs, following ref.~\cite{Wen}.

The simplest filling fraction where the paired spin-singlet 
states that we propose are possible is $\nu=\frac{2}{3}$. At 
that same filling fraction, there exists an abelian spin-singlet 
state, described by composite fermions with  
anti-parallel flux attachment \cite{WDJ}. To distinguish the
different states, one may consider the exponents for 
various tunneling processes. For the paired spin-singlet state the 
scaling dimensions for electrons, holons and spinons are 
$g_{\rm el} = m+2$, $g_{\rm hol} = \frac{2m+5}{16m+8}$, and
$g_{\rm sp} = \frac{5}{8}$, respectively. Thus, for tunneling 
through the bulk, the holon is the most relevant particle 
(for $m\geq 1$), while the $I-V$ for tunneling electrons from a 
Fermi-liquid into the edge is $I \sim V^{g_e} = V^{m+2}$.
According to \cite{LF2}, the scaling dimensions for the
composite fermion spin-singlet state at $\nu=\frac{2}{3}$ are
$g_{\rm el} = 2$, $g_{\rm qp} = \frac{2}{3}$. They give rise to a
quadratic $I-V$ for electron tunneling, in contrast to the cubic
$I-V$ for the paired state.
Another way to distinguish the two states is via the spin-Hall conductance,
which has opposite sign as compared to the ordinary Hall conductance for
the abelian state. For the paired spin-singlet state both conductances have
the same sign.  

There are two ways in which the paired state Eq.~(\ref{eq:paired})
can be relevant in a double-layer geometry. First, as already mentioned,
there is the possibility of a transition from a double-layer state for 
spin-full electrons, Eq.~(\ref{eq:so6}), into a single-layer 
paired state. A second possibility is a realization of the paired
state as a double-layer state for spin-polarized electrons, with 
the layer index playing the role of the spin-index.

As is the case for the pfaffian and the NASS states, these states can
be generalized to states which show clustering instead of pairing. 
Starting from an $SO(5)_k$ symmetry structure, one derives states
that allow clusters of up to $2k$ particles of equal spin, with
filling fractions given by $\nu=\frac{2k}{2km+1}$.

This research is supported in part by the 
Foundation FOM of the Netherlands and by the Netherlands
Organization for Scientific Research (NWO). A.W.W.L.\ acknowledges
the Institute for Theoretical Physics at the University of 
Amsterdam for hospitality. His research is supported by NSF under
Grant DMR-00-75064.


\vskip 3mm

\begin{thebibliography}{99}

\bibitem{And}
P.~W.~Anderson, Science {\bf 235}, 1196 (1987).

\bibitem{spch}
For a recent discussion, see
T.~Senthil and M.~P.~A.~Fisher, 
J.\ Phys.\ {\bf A 34}, L119 (2001).

\bibitem{WG}
N.~K.~Wilkin and J.~M.~F.~Gunn,
Phys.\ Rev.\ Lett.\ {\bf 84}, 6 (2000).

\bibitem{MR} 
G.~Moore and N.~Read, Nucl.\ Phys.\ {\bf B360}, 362 (1991).

\bibitem{BF} 
A.~V.~Balatsky and E.~Fradkin, Phys.\ Rev.\ {\bf B43},
10622 (1992).

\bibitem{MiR}
M.~Milovanovic and N.~Read, Phys.\ Rev.\ 
{\bf B56}, 1461 (1997).

\bibitem{Eis}
For a review, see J.~P.~Eisenstein in: {\it Perspectives in
Quantum Hall Effects}, S.~Das Sarma and A.~Pinczuk editors
(Wiley, New York 1997). 

\bibitem{Kang}
W.~Kang {\it et al.}, Phys.\ Rev.\ {\bf B56}, 12776 (1997);
H.~Cho {\it et al.}, Phys.\ Rev.\ Lett.\ {\bf 81}, 2522 (1998).

\bibitem{Halp}
B.~Halperin, Helv.\ Phys.\ Acta\ {\bf 56}, 75 (1983).

\bibitem{Rez}
E.~H.~Rezayi, Phys.\ Rev.\ {\bf B39}, 13541 (1989).

\bibitem{WDJ}
X.~G.~Wu, G.~Dev and J.~K.~Jain,
Phys.\ Rev.\ Lett.\ {\bf 71}, 153 (1993).

\bibitem{LF}
A.~Lopez and E.~Fradkin, Phys.\ Rev.\ {\bf B51},
4347 (1995).

\bibitem{Read}
N.~Read, Physica {\bf B298}, 121 (2001).

\bibitem{ARRS}
E.~Ardonne and K.~Schoutens, 
Phys.\ Rev.\ Lett.\ {\bf 82}, 5096 (1999);
E.~Ardonne, N.~Read, E.~Rezayi and K.~Schoutens,
Nucl.\ Phys.\ {\bf B607}, 549 (2001).

\bibitem{Rezpri}
E.~Rezayi, private communication.

\bibitem{Laugh}
R.~B.~Laughlin,
Phys.\ Rev.\ Lett.\ {\bf 50}, 1395 (1983).

\bibitem{SF}
T.~Senthil and M.~P.~A.~Fisher,
Phys.\ Rev.\ {\bf B62}, 7850 (2000).

\bibitem{RR} 
N.~Read and E.~Rezayi, 
Phys.\ Rev.\ {\bf B54}, 16864 (1996).

\bibitem{NW} 
C.~Nayak and F.~Wilczek, Nucl.\ Phys.\ {\bf B479}, 529 (1996).

\bibitem{RG}
N.~Read and D.~Green,
Phys.\ Rev.\ {\bf B61}, 10267 (2000).

\bibitem{Wen}
X.-G.~Wen, Phys.\ Rev.\ Lett.\ {\bf 84}, 3950 (2000).

\bibitem{Wen1}
X.-G.~Wen, Adv.\ Phys.\ {\bf 44}, 405 (1995).

\bibitem{ABGS}
E.~Ardonne, P.~Bouwknegt, S.~Guruswamy and K.~Schoutens,
Phys.\ Rev.\ {\bf B61}, 10298 (2000);
E.~Ardonne, P.~Bouwknegt and K.~Schoutens,
J.\ Stat.\ Phys. {\bf 102}, 421 (2001).

\bibitem{GSBCR}
S.~Guruswamy and K.~Schoutens,
Nucl.\ Phys.\ {\bf B556}, 530 (1999); 
P.~Bouwknegt, L.-H.~Chim and D.~Ridout,
Nucl.\ Phys.\ {\bf B572}, 547 (2000).

\bibitem{LF2}
A.~Lopez and E.~Fradkin, Phys.\ Rev.\ {\bf B63},
085306 (2001); Phys.\ Rev.\ {\bf B64}, 
049903(E) (2001).

\end{thebibliography}
\end{document}